\documentclass[12pt]{iopart}
\usepackage{amsfonts}
\usepackage{amssymb}
\usepackage{epsfig}
\usepackage{graphicx}

\newcommand{\pdif}[2]{\frac{\partial #1}{\partial #2}}

\newcommand{\new}{\nonumber\\}
\begin{document}

\title{The decoupling of the glass transitions in the two-component $p$-spin spherical model}

\author{Harukuni Ikeda$^1$, Atsushi Ikeda$^2$}
\address{
$^1$ Department of Physics, Nagoya University - Nagoya, 464-8602, Japan
$^2$ Fukui Institute for Fundamental Chemistry, Kyoto University - Takano-Nishihiraki-cho 34-4,
Sakyo-ku, Kyoto, 606-8103, Japan}
\ead{atsushi.ikeda@fukui.kyoto-u.ac.jp}

\begin{abstract}
Binary mixtures of large and small particles with disparate size ratio
exhibit a rich phenomenology at their glass transition points.  In order to gain
insights on such systems, we introduce and study a two-component
version of the $p$-spin spherical spin glass model.  We employ the replica method
to calculate the free energy and the phase diagram.  We show that when the strengths
of the interactions of each component are not widely separated, the model
has only one glass phase characterized by the conventional one-step
replica symmetry breaking.  However when the strengths of the interactions
are well separated, the model has three glass phases depending on temperature and component ratio. 
One is the ``single'' glass phase in which only the spins of one component are frozen 
while the spins of the other component remain mobile. 
This phase is characterized by the one-step replica symmetry breaking. 
The second is the ``double'' glass phase obtained by cooling further the single glass phase, 
in which the spins of the remaining mobile component are also frozen. 
This phase is characterized by the two-step replica symmetry breaking.  
The third is also the ``double'' glass phase, which however is formed 
by the simultaneous freezing of the spins of both components at the same temperatures 
and is characterized by the one-step replica symmetry breaking. 
We discuss the implications of these results 
for the glass transitions of binary mixtures. 
\end{abstract}

\maketitle
\section{Introduction}

The $p$-spin spherical model (PSM) has been playing important roles in
the study of the glass transition of liquids, because it shares
many common properties in dynamics and thermodynamics with glass
forming liquids~\cite{KT1,KT2,BB,Cavagna}.  The PSM is the infinite range
spin glass model in which soft spins interact through $p$-body interactions
with random quenched couplings~\cite{SGpede}.  
The dynamics of
the PSM can be solved semi-analytically~\cite{KT1,SGpede,CS2}.
Particularly at $p=3$, the time correlation function is known to obey the dynamical
equation mathematically equivalent with the mode-coupling theory
(MCT) equation of the glass transition~\cite{KT1,Gotze}.  The system is
ergodic at high temperature, however as temperature is decreased,  
the time correlation function exhibits the two step relaxation behavior and 
the relaxation becomes slower and slower.
Eventually the relaxation time diverges and the spins get frozen, 
which is called the {\it dynamical transition}. 
Also the thermodynamics of the PSM can be solved semi-analytically by the
replica method with the one-step replica symmetry breaking (1RSB)
ansatz~\cite{SGpede,CS1}.  
As temperature is lowered from above, 
the phase space of the system in the paramagnetic state splits into many metastable glassy states 
exactly at the dynamical transition temperature.  
As the system is cooled further, the logarithm of the
number of these states or the complexity, which corresponds to the configurational entropy
in glass forming liquids, decreases and
eventually becomes zero where the {\it thermodynamic glass
transition} takes place. In the glass phase, the free energy of the model is 
dominated by the lower energy states.  
The similarity between the PSM and
glass forming liquids has many to believe that they are in the same
class of random glassy systems, at least in the mean-field limit~\cite{BB,Cavagna}.

However, real glass formers often exhibit richer and more anomalous dynamical behaviors, 
all of which can not be captured by the PSM.  
In this work, we particularly focus on the ``decoupling'' phenomenon 
often observed in multi-component glass formers.  
This is the phenomenon in which the slowing down of the dynamics of each component 
occurs separately at different regions of the densities and the temperatures, 
hence some components are frozen into a glass state while the others remain mobile.  
There is a wide variety of materials showing the decoupling phenomenon, such as ionic glasses 
and metallic glasses~\cite{Angell}. 
The simplest example among them is presumably 
a binary mixture of large and small particles with disparate size ratio~\cite{
Dhont1, Dhont2, Pham1, Eckert, Pham2, Sentjabrskaja1, Sentjabrskaja2,
Hendricks, Moreno1, Moreno2, VH, Bosse, Voigtmann, Amokrane1}.  
When the size ratio is sufficiently large, it is observed in
experiments~\cite{Dhont1, Dhont2} and
simulations~\cite{Moreno1,Moreno2,VH} that there are two distinct glass phases in this model: 
the ``single'' glass where only large
particles are arrested while small particles are left mobile, and 
the ``double'' glass where both small and
large particles are arrested. 
Despite of simplicity of the model, this decoupling phenomenon of binary mixtures
is not fully understood theoretically. 
It is encouraging that the MCT can 
predict this behavior qualitatively~\cite{Bosse, Voigtmann, Amokrane1}. 
However the MCT is derived using numerous uncontrorable approximations, 
which are not guaranteed to be valid for binary mixtures with disparate size ratio.  
Even for monodisperse systems, there is an argument whether or not 
the MCT is a true ``mean-field theory'' to describe the dynamics of the glass transition~\cite{
Schilling, IM, Jacquin, Maimbourg}. 
Moreover the transition predicted by the MCT only exists in the 
mean-field limit and is washed away in finite dimensions~\cite{BB,Cavagna}.  

Can any of spin glass models qualitatively capture these rich
behaviors of the glass transitions of binary mixtures? 
If so, analysis of such models should facilitate the study of binary mixtures  
because spin models can be analyzed rigorously at least in the mean-field limit. 
Related to this point, Crisanti and Leuzzi generalized 
the PSM to include two distinct energy scales of the interactions~\cite{CL1, CL2, CL3, CL4}.  
They considered the $s+p$-spin spherical
model, where all spins interact through both $s$-body 
and $p$-body interactions. 
This model is potentially related to the glass forming liquids 
in which molecules are subject to two different types of interactions.  
They found that there is a variety of glass phases 
characterized by the series of replica symmetry breaking~\cite{CL3} and that 
the model exhibits rich dynamical behaviors 
such as three-step relaxation of the time correlation function~\cite{CL4}.
However to the best of our knowledge, 
there exists no study on the spin glass model which exhibits the single and double glass transitions 
and the decoupling of dynamics of one of the components from the other, 
as observed for binary mixtures.  

In this work, we extend the PSM  so as to mimic 
binary mixtures of particles with disparate size ratio.  
Our model is a two component version of the PSM, which is
composed of weakly interacting spins (weak spins) and strongly
interacting spins (strong spins).  We employ the replica theory to study
the thermodynamics of the model.  We found that the model has the glass
phases characterized by either conventional 1RSB 
and the two-step replica symmetry breaking (2RSB). We show that 
the interplay between the 1RSB and the 2RSB solutions 
results in the decoupling of the glass transitions of weak spins from that of strong spins.  
We also show that our two component PSM is directly related to the
randomly pinned PSM, which has been studied recently~\cite{CammarotaJCP}.  
Finally based on the results, we discuss the validity of the predictions of the MCT 
for the multiple glass phases of binary mixtures. 

The organization of the paper is as follows.  In Section~II, we introduce
the model.  In Section~III, we use the replica theory to express the free
energy in terms of the spin glass order parameters.  In Section~IV, by 
numerical minimization of the free energy, we obtain the temperature
evolutions of the order parameters, the phase diagrams, and the
thermodynamic quantities of the model.  In Sections~V and VI,
we discuss the results and conclude the work.

\section{Model}

We consider a two component version of the PSM with
$p=3$.  The model is composed of $N_1$ spins of the component 1 and
$N_2$ spins of the component 2, with $N = N_1 + N_2$.  The spin
variables for each component are denoted as $\sigma_{1,i}~(i = 1, \cdots, N_1)$
and $\sigma_{2,i}~(i = 1, \cdots, N_2)$, respectively. 
They obey the spherical
constraints $N_1 = \sum_i \sigma_{1, i}^2$ and $N_2 = \sum_i
\sigma_{2, i}^2$.  The Hamiltonian of the model is
\begin{eqnarray}
H = \sum_{\alpha, \beta, \gamma = 1, 2} \sum_{i_{\alpha}, j_{\beta}, k_{\gamma}} 
J^{(\alpha \beta \gamma)}_{i_{\alpha} j_{\beta} k_{\gamma}} 
\sigma_{\alpha, i_{\alpha}} \sigma_{\beta, j_{\beta}} \sigma_{\gamma, k_{\gamma}},  
\end{eqnarray}
where the greek indices are used to indicate components, 
the roman indices are for spins, 
and $J^{(\alpha \beta \gamma)}_{i_{\alpha} j_{\beta} k_{\gamma}}$ is the coupling constant among the three spins, 
which is the Gaussian random variables with zero mean. 
In order to render the analysis tractable, we consider the case 
where $J^{(\alpha \beta \gamma)}_{i_{\alpha} j_{\beta} k_{\gamma}}$ is characterized by only the two values, i.e.
\begin{eqnarray}
\overline{(J^{(\alpha \beta \gamma)})^2} = 
\cases{
\frac {3 J_1^2}{N^2} & $(\alpha, \beta ,\gamma) = (1, 1 ,1)$,\\ 
\frac {3 J_2^2}{N^2} & $(\alpha, \beta ,\gamma) \neq (1, 1 ,1)$.} 
\end{eqnarray}
Here $J_1$ and $J_2$ are the typical energy scales of the interactions of the
component 1 and 2, respectively.  We set $J_1 > J_2$, hence the
component 1 is ``strong'' spins and 2 is ``weak'' spins.  The control
parameters of the model are the ratio of the strengths of the
interactions $J = J_2/J_1$, the fraction of strong spins $c =
N_1/N$, and the temperature $T$.  We use $J_1$, $J_1/k_B$ and $k_B$ 
for the units of the energy, temperature, and the entropy, respectively,  
where $k_B$ is the Boltzmann constant.
All the results are obtained in the
thermodynamic limit. 

\section{Free energy calculation}

We calculate the free energy of the model using the standard replica
method.  In the method, the free energy
of the original model is obtained by taking the limit $n \to 0$ of the free energy of $n$ replicas.
The procedure of this calculation for the one-component PSM is
well documented in Ref.~\cite{SGpede, CS1}.  Following the same procedure, we 
write down the free energy of the two-component PSM as 
\begin{eqnarray}
- F/T = \lim_{n \to 0} \frac{1}{2 n} \max G_n(\mat{Q}, \mat{P}) \label{fstart}
\end{eqnarray}
with 
\begin{eqnarray}
\fl
G_n = \sum_{ab} \frac{1}{2T^2} 
\Bigl[ (c Q_{ab})^3 + 3 J^2 (c Q_{ab})^2 ((1-c) P_{ab}) + 3 J^2 (c Q_{ab}) ((1-c) P_{ab})^2 + J^2 ((1-c) P_{ab})^3 \Bigr] \nonumber \\
+ c \log \det \mat{Q}  + (1-c) \log \det \mat{P} + n (1 + \log 2\pi), \label{gn}
\end{eqnarray}
where $\mat{Q}$ and $\mat{P}$ denote the overlap matrices for the component 1 and 2, 
each component of which is defined by 
$Q_{ab} = \frac{1}{N_1} \sum_i \sigma_{1, i}^{(a)}
\sigma_{1, i}^{(b)}$ and $P_{ab} = \frac{1}{N_2} \sum_i \sigma_{2, i}^{(a)}
\sigma_{2, i}^{(b)}$. 
$\max G_n(\mat{Q}, \mat{P})$ means that the function $G_n$ is
maximized with respect to the matrices $\mat{Q}$ and $\mat{P}$.

\subsection{1RSB ansatz}

\begin{figure}
\centering \includegraphics[width=8.5cm]{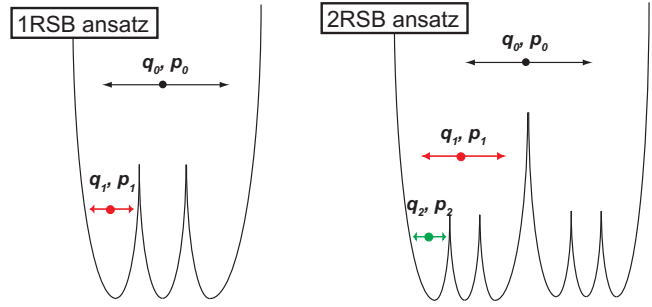} \caption{\label{fig1}
Sketch of the free energy landscapes corresponding to the 1RSB solution (left) and the 2RSB solution (right).  
There is the intermediate level of the hierarchy of states in the 2RSB solution.  }
\end{figure}

In the case of the one-component PSM, it is known that the 1RSB ansatz
gives the correct solution. The 1RSB ansatz assumes that the overlap matrices have a
one-step hierarchical structure.  In our model, this ansatz reads explicitly 
\numparts
\begin{eqnarray}
Q_{ab} =  (1-q_1) \delta_{ab} + (q_1 - q_0) \epsilon^{m_1}_{ab} + q_0, \label{o1rsba} \\ 
P_{ab} =  (1-p_1) \delta_{ab} + (p_1 - p_0) \epsilon^{m_1}_{ab} + p_0, \label{o1rsbb}
\end{eqnarray}
\endnumparts
where $\delta_{ab}$ is the Kronecker delta and 
\begin{eqnarray}
\epsilon_{ab}^{m_1} = 
\cases{
1 & if \ $a$ \ and \ $b$ \ are in a diagnal block of \ $m_1 \times m_1$\\
0 & otherwise}.
\end{eqnarray}
Here, $q_1$ and $p_1$ are called the self overlaps,  
which are the overlaps within the same glassy states, and 
$q_0$ and $p_0$ are the overlaps between different glassy
states (Figure~\ref{fig1} left). We can assume $q_0 = p_0 =
0$ that is valid for the PSM without external fields. Substituting 
equations~(\ref{o1rsba}) and (\ref{o1rsbb}) into 
equation~(\ref{gn}), and taking the limit $n \to 0$ in 
equation~(\ref{fstart}), we obtain
\begin{eqnarray}
- F/T = \frac{1}{2} (1 + \log 2\pi) + x_1 + x_2 + x_3 + x_4 +
\min_{m_1,q_1,p_1} G_{1RSB} 
\end{eqnarray}
with 
\begin{eqnarray}
\fl
G_{1RSB} = (m_1 - 1)[ x_1 q_1^3 + x_2 q_1^2 p_1 + x_3q_1 p_1^2 + x_4 p_1^3 ] \nonumber \\
\fl
+ \frac{c}{2} \Bigl[ \log (1 - q_1) + \frac{1}{m_1} \log \frac{1 + (m_1-1) q_1}{1 - q_1} \Bigr] 
+ \frac{1-c}{2} \Bigl[ \log (1 - p_1) + \frac{1}{m_1} \log \frac{1 + (m_1-1) p_1}{1 - p_1} \Bigr], \nonumber \\
\label{gform1}
\end{eqnarray}
where $x_1 = c^3/4T^2$, $x_2 = 3 c^2 (1-c) J^2/4T^2$, $x_3 = 3 c (1-c)^2 J^2/4T^2$, 
and $x_4 = (1-c)^3  J^2/4T^2$.  
The breaking parameter $m_1$ should be limited to $0 \leq m_1 \leq 1$ in the limit $n \to 0$.  
When $m_1 = 1$, this 1RSB free energy reduces to that of the paramagnetic state.
When $G_{1RSB}$ is
extremized with respect to $q_1$, $p_1$, and $m_1$, 
the 1RSB solution of the model is obtained.  
The 1RSB dynamical transition is defined as 
the transition where the overlaps $q_1$ and $p_1$ change discontinuously, 
and the 1RSB thermodynamic transition is defined as 
the transition where the 1RSB solution with $m_1 \neq 1$ becomes stable. 

\subsection{2RSB ansatz}

Our model is the two component PSM with the two distinct energy
scales $J_1$ and $J_2$, which make the thermodynamic phase diagram more complex. 
Especially there is no guarantee that the 1RSB ansatz gives the stable solution. 
Therefore, we have to allow the two-step hierarchical
structure of the overlap matrices: \numparts
\begin{eqnarray}
Q_{ab} =  (1-q_2) \delta_{ab} + (q_2 - q_1) \epsilon^{m_2}_{ab} + (q_1 - q_0) \epsilon^{m_1}_{ab} + q_0, \label{o2rsba} \\
P_{ab} =  (1-p_2) \delta_{ab} + (p_2 - p_1) \epsilon^{m_2}_{ab} + (p_1 - p_0) \epsilon^{m_1}_{ab} + p_0, \label{o2rsbb} 
\end{eqnarray}
\endnumparts which are called the 2RSB ansatz. 
This ansatz corresponds 
to the two-step hierarchical structure of the free energy landscape
as depicted schematically in Figure~\ref{fig1} right.  
Here, $q_2$ and $p_2$ are the self overlaps, 
$q_1$ and $p_1$ are the overlaps between the different glassy
states in the same group in the intermediate level of the hierarchy, 
and $q_0$ and $p_0$ are the overlaps between the different glassy 
states in the different groups. 
Substituting equations~(\ref{o2rsba}) and (\ref{o2rsbb}) into
equation~(\ref{gn}) and taking the limit $n \to 0$
in equation~(\ref{fstart}), we obtain
\begin{eqnarray}
- F/T = \frac{1}{2} (1 + \log 2\pi) + x_1 + x_2 + x_3 + x_4 + \min_{m_1,m_2,q_1,q_2,p_1,p_2} G_{2RSB} 
\end{eqnarray}
with 
\begin{eqnarray}
\fl
G_{2RSB} = (m_2 -1) [ x_1 q_2^3 + x_2 q_2^2 p_2 + x_3 q_2 p_2^2 + x_4 p_2^3] 
+ (m_1 - m_2) [x_1 q_1^3 + x_2 q_1^2 p_1 + x_3 q_1 p_1^2 + x_4 p_1^3 ] \nonumber \\
\fl
+ \frac{c}{2} \Bigl[
\log (1 - q_2) + \frac{1}{m_1} \log \frac{1 +(m_2-1) q_2 + (m_1-m_2) q_1}{1 +(m_2-1)q_2 - m_2 q_1} + \frac{1}{m_2} \log \frac{1 +(m_2-1)q_2 - m_2 q_1}{1 - q_2} \Bigr] \nonumber \\ 
\fl
+ \frac{1-c}{2} \Bigl[
\log (1 - p_2) + \frac{1}{m_1} \log \frac{1 +(m_2-1) p_2 + (m_1-m_2) p_1}{1 +(m_2-1)p_2 - m_2 p_1} + \frac{1}{m_2} \log \frac{1 +(m_2-1)p_2 - m_2 p_1}{1 - p_2} \Bigr]. \nonumber \\ 
\label{gform2}
\end{eqnarray}
The breaking parameters $m_1$ and $m_2$
should be limited to $0 \leq m_1 < m_2 \leq 1$. 
When $m_2 = 1$, the 2RSB free energy $G_{2RSB}$, equation (\ref{gform2}), 
reduces to the 1RSB free energy $G_{1RSB}$, equation (\ref{gform1}). 
By minimizing $G_{2RSB}$ in equation~(\ref{gform2}) with respect to
the order parameters $q_1$, $p_1$, $q_2$, $p_2$, $m_1$ and $m_2$, the
free energy and the order parameters of the original system is obtained
within the 2RSB ansatz.  
The 2RSB dynamical transition is defined as 
the transition where the overlaps $q_2$ and $p_2$ change discontinuously, 
and the 2RSB thermodynamic transition is defined as 
the transition where the 2RSB solution with $m_2 \neq 1$ becomes stable~\footnote{ 
We did not explore the possibilities of RSB of the higher order than 2RSB.  
Note that the 2RSB is guaranteed to be sufficient at least in the limit $J \to 0$ 
because this limit corresponds to the randomly pinned PSM 
where the solution corresponding to the 2RSB is verified to be stable~\cite{CammarotaJCP}.}. 

\subsection{Numerical minimization of $G_{2RSB}$}

We focus on the minimization of $G_{2RSB}$. 
We do not need to analyze $G_{1RSB}$ separately, 
because $G_{1RSB}$ is included in $G_{2RSB}$ as discussed above.  
We employ the following
numerical method to minimize $G_{2RSB}$.  For a given $c$, we first focus
on a low enough temperature (in practice, we set $T = J/3$) 
and minimize $G_{2RSB}$ by the steepest descent method.  
We take a
special care in this procedure
because the calculation easily gets stuck in locally stable solutions. 
In order to avoid this unwanted effect, 
we first slice the $(m_1,m_2)$ space to 50 grid points and minimize $G_{2RSB}$ with respect to
$q_1$, $p_1$, $q_2$ and $p_2$ at each grid point.  We seek for the set 
of $(m_1,m_2)$ which minimizes $G_{2RSB}$.  
Using this $(m_1,m_2)$ as an initial guess, we perform
the full steepest descent optimization of all the order parameters.
After obtaining the optimized solution at the lowest temperature, we gradually
increase the temperature and employ the steepest descent method to
minimize $G_{2RSB}$ at each temperature, using the optimal values of the
order parameters at the lower temperature as an initial guess.

\section{Phase diagrams and thermodynamic quantities} 

In this section, we show the phase diagrams and the thermodynamic
quantities of the two-component PSM obtained by the minimization of
$G_{2RSB}$.  We find that the model has a variety of glass phases
including the ``single'' and the ``double'' glasses when $J$ is very different from 1. 

\subsection{Order parameters and phase diagrams}

\subsubsection{$J=0.3$.}

\begin{figure}
\centering
\includegraphics[width=8.5cm]{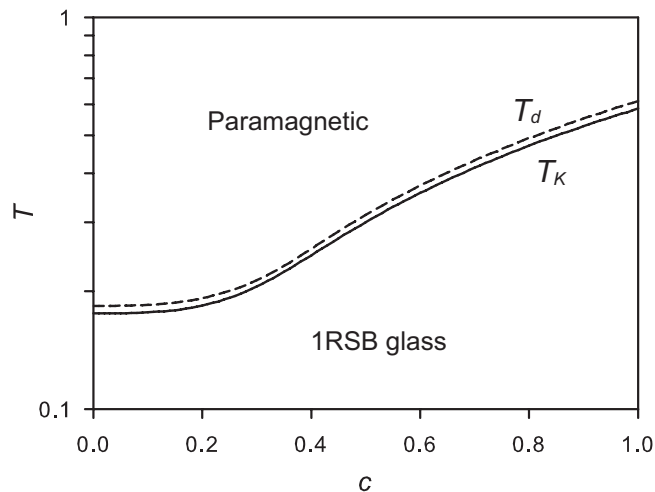}
\caption{\label{fig2} 
The phase diagram of the two-component PSM at $J=0.3$. 
$T_d$ (dashed line) is the 1RSB dynamical transition temperatures; 
$T_K$ (solid line) is the 1RSB thermodynamic transition temperatures.
There is only one glass phase characterized by the 1RSB solution at $J=0.3$. 
}
\end{figure}

We start with $J=0.3$, 
which is not very different from 1. 
We show the phase diagram in Figure~\ref{fig2}.  
There are only the paramagnetic phase and the 1RSB glass phase. 
The two phases are separated by the 1RSB thermodynamic transition line $T_K(c)$. 
The 1RSB dynamical transition line $T_d(c)$ is located at slightly higher temperatures. 
Note that $T_d(c)$ and $T_K(c)$ for $c=1$ match with the results of 
the one-component PSM of the strong spins. 
They are $T_d(c=1) = 0.612$ and $T_K(c=1) = 0.586$. 
$T_d(c)$ and $T_K(c)$ for $c=0$ are identical to those for $c=1$ 
aside from the obvious factor of $J$, 
which defines the unit of the energy, i.e.,  
$T_d(c=0) = 0.612J = 0.184$ and $T_K(c=0) = 0.586 J = 0.176$.  
The transition lines smoothly connect these two limiting cases.

\begin{figure}
\centering \includegraphics[width=14cm]{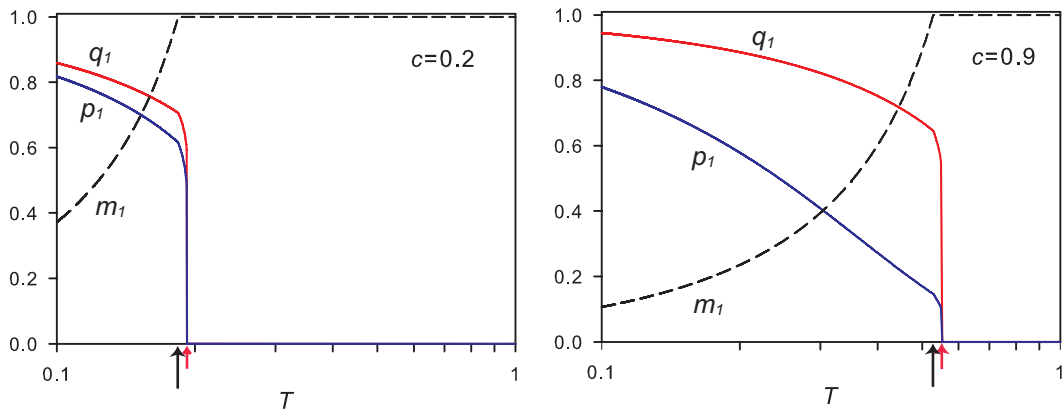} \caption{\label{fig3}
Temperature dependence of the overlaps $q_1$ and $p_1$ and the breaking parameter $m_1$ at $c=0.2$ (left) and 0.9 (right) at $J=0.3$.
The short red and long black arrows indicate 
the 1RSB dynamical and thermodynamic transition temperatures $T_d$ and $T_K$, respectively. 
}
\end{figure}

In order to gain more insights, in Figure~\ref{fig3}, we plot the temperature dependence of the
optimized overlaps $q_1$ and $p_1$ and the breaking parameter $m_1$ 
at two representative values of $c=0.2$ and 0.9. 
At $c=0.2$ (Figure~\ref{fig3} left), as temperature is decreased, 
the overlaps $q_1$ and $p_1$ jump from zero 
while the breaking parameter remains constant $m_1 = 1$
at the 1RSB dynamical transition temperature $T_d$.  
$m_1$ suddenly starts decreasing from 1 
at the 1RSB thermodynamic transition temperature $T_K$. 
The 1RSB dynamical and thermodynamic transition temperatures ($T_d(c=0.2)=0.192$ and $T_K(c=0.2)=0.184$) 
are close to those of the one-component PSM of weak spins ($T_d(c=0)=0.184$ and $T_K(c=0)=0.176$), 
indicating that the 1RSB transition at $c=0.2$  
is driven mainly by the freezing of the weak spins. 
Note that both the values of $q_1$ and $p_1$ 
just below the transition temperatures are larger than 0.5 and are close to each other, 
which can be interpreted that both the strong and weak spins are frozen equally strongly at this transition. 

Behaviors at $c=0.9$ (Figure~\ref{fig3} right) are qualitatively similar to those at $c=0.2$.  
The only differences are that 
(i) the 1RSB dynamical and thermodynamic transition temperatures  
($T_d(c=0.9) = 0.552$ and $T_K(c=0.9) = 0.528$) are close to those  
of the one-component PSM of the {\it strong} spins ($T_d(c=1) = 0.612$ and $T_K(c=1) = 0.586$) 
and that (ii) the value of $q_1$ is larger than 0.5 while the value of $p_1$ is much smaller just below the transition temperatures. 
These results can be interpreted that the 1RSB transition at $c=0.9$  
is driven mainly by the freezing of the strong spins, and 
the weak spins are not strongly frozen at this transition. 
However we emphasize that this difference is only quantitative 
and the overlaps of the weak and strong spins vary continuously as $c$ changes. 
Namely there is no clear signature of the decoupling of 
the glass transitions of the weak and strong spins. 

\subsubsection{$J=0.03$.}

\begin{figure}
\centering
\includegraphics[width=14cm]{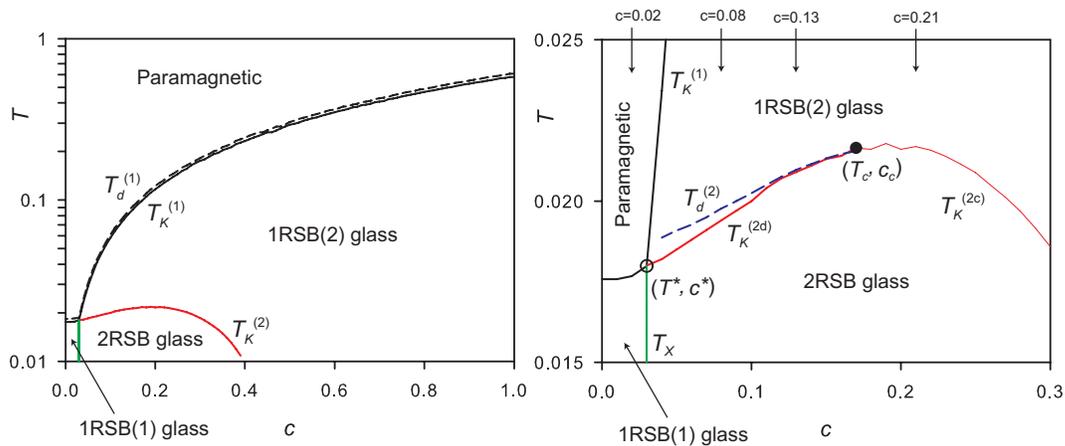}
\caption{\label{fig4} 
The phase diagram at $J=0.03$. 
(left) Overall view. 
There are the paramagnetic phase and the three glass phases, the 1RSB(1), the 1RSB(2), and the 2RSB. 
$T_d^{(1)}$ and $T_K^{(1)}$ are the 1RSB dynamical and thermodynamic transition temperatures, respectively. 
$T_K^{(2)}$ is the 2RSB thermodynamic transition temperature. 
(right) Zoom on the the 2RSB glass region. 
$T_K^{(2)}$ is composed of the discontinuous and continuous 2RSB transition temperatures, $T_K^{(2d)}$ and $T_K^{(2c)}$. 
$T_d^{(2)}$ is the 2RSB dynamical transition temperature, 
which terminates at the critical point $(T_c,c_c) \approx (0.022,0.17)$. 
$T_X$ is the phase boundary between the 1RSB(1) and the 2RSB glass phases. 
The three thermodynamic transition lines, $T_K^{(1)}$,  $T_K^{(2d)}$, and $T_X$, meet 
at $(T^{\star},c^{\star}) \approx (0.018,0.03)$. 
The four downwards arrows indicate the four representative values of $c$, 
for which the temperature evolutions of the overlaps and the thermodynamic quantities are presented in Figures~\ref{fig5} and \ref{fig7}. 
}
\end{figure}

Next we focus on $J=0.03$, which is much smaller than 1. 
In Figure~\ref{fig4} left, we show the phase diagram, 
which is qualitatively different from that at $J=0.3$.  
One finds that there are three glass phases.  
We refer to them as the 1RSB(1), the 1RSB(2), and the 2RSB glass phases.  
The paramagnetic phase is separated from the 1RSB(1) 
and the 1RSB(2) glass phases by the 1RSB thermodynamic transition line $T_K^{(1)}(c)$.  
The associated 1RSB dynamical transition line $T_d^{(1)}(c)$ is located at slightly above $T_K^{(1)}(c)$. 
The 2RSB glass phase is located below the 1RSB(2) glass phase. 
The 1RSB(2) and the 2RSB glass phases are separated 
by the 2RSB thermodynamic transition line $T_K^{(2)}(c)$. 
To present the details of the 2RSB glass phase region, 
we show the zoom in Figure~\ref{fig4} right. 
The 2RSB thermodynamic transition line $T_K^{(2)}(c)$ is 
composed of two parts: 
$T_K^{(2d)}(c)$ at lower fraction of strong spins 
and $T_K^{(2c)}(c)$ at higher fraction of strong spins, 
depending on the discontinuous and continuous nature of the transition across this temperature.  
The 2RSB dynamical transition line $T_d^{(2)}(c)$ is located at slightly above $T_K^{(2d)}(c)$ 
and it terminates at the critical point $(T_c,c_c) \approx (0.022,0.17)$, 
at which the three transition lines $T_d^{(2)}(c)$, $T_K^{(2d)}(c)$ and $T_K^{(2c)}(c)$ meet. 
The thermodynamic transition line which separates 
the 1RSB(1) glass from the 2RSB glass is denoted as $T_X(c)$. 
The three thermodynamic transition lines, $T_K^{(1)}(c)$, $T_K^{(2d)}(c)$ and $T_X(c)$, 
meet at the point $(T^{\star},c^{\star}) \approx (0.018,0.03)$, 
which is the meeting point of all the four phases.   
As $c$ increases, the 1RSB thermodynamic transition line $T_K^{(1)}(c)$ 
sharply bends upward at this point 
and the transition into the 1RSB(1) glass becomes the transition into the 1RSB(2) glass. 

\begin{figure}
\centering
\includegraphics[width=14cm]{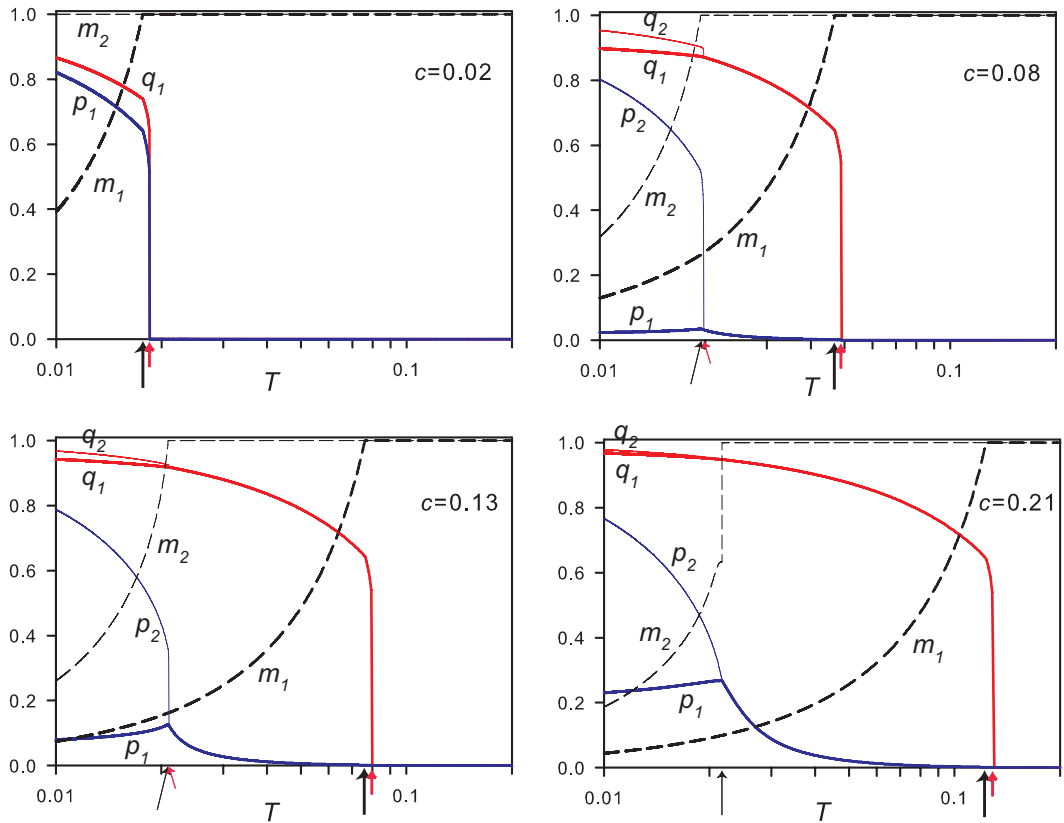}
\caption{\label{fig5} 
Temperature dependence of the overlaps, $q_1$, $p_1$, $q_2$, and $p_2$, and the breaking parameters, $m_1$, and $m_2$, 
at $c=0.02$ (upper left), 0.08 (upper right), 0.13 (lower left) and 0.21 (lower right) at $J=0.03$. 
The bold short red and long black arrows indicate 
the 1RSB dynamical and thermodynamic transition temperatures $T_d$ and $T_K$, respectively;  
the thin short red and long black arrows indicate 
the 2RSB dynamical and thermodynamic transition temperatures $T_d^{(2)}$ and $T_K^{(2d)}$ 
($T_K^{(2c)}$ for $c=0.21$), respectively, 
}
\end{figure}

In order to understand the nature of these phases,  
we plot the temperature dependence of the overlaps and the breaking parameters 
at four representative values of $c = 0.02$, 0.08, 0.13, and 0.21, in Figure~\ref{fig5}.  
These values of $c$ are indicated as arrows in the phase diagram, see Figure~\ref{fig4} right. 
We first focus on $c=0.02$ (Figure~\ref{fig5} upper left).  
When the temperature is decreased from above, 
$q_1$ and $p_1$ change discontinuously from zero 
at the 1RSB dynamical transition temperature $T_d^{(1)}$. 
While $m_2$ remains to be unity, $m_1$ suddenly starts decreasing from 1 at $T_K^{(1)}$, 
where the 1RSB thermodynamic transition from the paramagnetic phase to the 1RSB(1) glass phase takes place. 
Note that the 1RSB dynamical and thermodynamic transition temperatures 
($T_d^{(1)}(c=0.02)=0.0185$ and $T_K^{(1)}(c=0.02)=0.0177$) 
are close to those of the one-component PSM of the weak spins 
($T_d^{(1)}(c=0)=0.0184$ and $T_K^{(1)}(c=0)=0.0176$), 
which indicates that the transition into the 1RSB(1) glass phase is driven mainly 
by the freezing of the weak spins. 
Both the values of $q_1$ and $p_1$ are larger than 0.5 and are close to each other below $T_K^{(1)}$, 
which can be interpreted that both the strong and weak spins are frozen equally strongly in the 1RSB(1) glass phase. 

The upper right panel of Figure~\ref{fig5} shows the results at $c=0.08$, 
where there are the 1RSB transition from the paramagnetic to the 1RSB(2) glass phase, 
and the 2RSB transition from the 1RSB(2) to the 2RSB glass phase. 
The 1RSB dynamical and thermodynamic transitions at $T_d^{(1)}$ and $T_K^{(1)}$ are qualitatively the same as those for $c=0.02$. 
The only difference is that the value of $q_1$ is larger than 0.5 while $p_1$ is very close to zero. 
This result can be interpreted that the strong spins are frozen while the weak spins are {\it not} frozen in the 1RSB(2) glass phase. 
As we decrease temperature further, the 2RSB transition takes place. 
First at $T_d^{(2)}$, the overlaps $q_2$ and $p_2$ change discontinuously 
while the breaking parameter remains constant $m_2 = 1$. 
At $T_K^{(2d)}$, $m_2$ suddenly starts decreasing from 1, 
where the 2RSB thermodynamic transition takes place. 
Note that $p_2$ is larger than 0.5 just below $T_K^{(2d)}$,
which can be interpreted that the weak spins are also frozen in the 2RSB glass phase. 

Behaviors at $c=0.13$ (Figure~\ref{fig5} lower left) are qualitatively similar to those at
$c=0.08$. 
The only difference is that the discontinuities of $q_2$ and
$p_2$ at the 2RSB dynamical transition are smaller than those for $c=0.08$. 
As $c$ increases, the discontinuities at the 2RSB dynamical transition become smaller, 
and eventually the jump of $q_2$ and $p_2$ disappear at $c=0.17$.  
The lower right panel of Figure~\ref{fig5} shows the results at $c=0.21$. 
The overlaps $q_2$ and $p_2$ change from $q_1$ and $p_1$ continuously 
at $T_K^{(2c)}$, where the continuous 2RSB thermodynamic transition takes place. 
Interestingly, the change of the breaking parameter $m_2$ at $T_K^{(2c)}$ is not continuous 
as in the case at $T_K^{(2d)}$ but discontinuous. 

In summary, the 1RSB(2) glass corresponds to the ``single'' glass where only the strong spins are frozen, 
and the 1RSB(1) and the 2RSB glasses correspond to the
``double'' glass where both the weak and strong spins are frozen simultaneously.  
We emphasize that there is a clear difference 
between these two ``double'' glasses, the 1RSB(1) and the 2RSB.  
The transition into the 1RSB(1) glass phase is the simultaneous arrest of the weak and the strong spins.  
This transition is mainly driven by the freezing of the weak spins. 
On the other hand, 
the transition into the 2RSB glass phase is the arrest of the weak spins 
in the presence of the frozen strong spins which already undergoes 
the glass transition at much higher temperature. 
The difference becomes
clearer when one considers the free energy landscape of these phases.  In
the 1RSB(1) glass phase, the landscape can be characterized by the
one-step hierarchical structure (Figure~\ref{fig1} left).  Only the self
overlap $q_1$ and $p_1$ have large values, and $q_0$ and $p_0$ are zero. 
This means that the different
glassy states have completely different configurations of spins. 
On the other hand in the 2RSB glass phase, the
landscape has the two-step hierarchical structure (Figure~\ref{fig1}
right).  Not only the self overlaps $q_2$ and $p_2$, but also the overlap
$q_1$ have large values.  This means that 
several glassy states in the same group of the intermediate level of the hierarchy 
share almost the same configuration of the strong spins. 
In other words, the phase space is divided into a multi-valley structure 
 corresponding to configurations of the strong spins and each of the valley is divided 
into a subgroups of multi-valley structure corresponding to configurations of the weak spins.
Note that the 1RSB(1) and the 2RSB glasses are separated 
by the glass-glass transition $T_X$ as shown in Figure~\ref{fig4}. 
As one crosses $T_X$, the overlaps discontinuously change (not shown).  


Finally we obtain a semi-analytical expression for the continuous 2RSB transition temperature $T_K^{(2c)}$. 
This is possible because $q_2-q_1$ and $p_2 - p_1$ are small just below $T_K^{(2c)}$ 
and thus the 2RSB solution can be expressed 
by the perturbation around the 1RSB solution (see Appendix for details). 
In Figure~\ref{figa1}, we compare $T_K^{(2c)}$ calculated by the perturbation theory 
with those calculated by the minimization of $G_{2RSB}$. 
The two results are almost identical, confirming that our numerical minimization of $G_{2RSB}$ is reliable. 

\subsubsection{The decoupling of the glass transitions of the weak and the strong spins.}

We showed that there is only one glass phase at $J=0.3$, 
whereas at $J=0.03$ 
the decoupling of the glass transitions of the weak and the strong spins takes place and, 
as a result, the three glass phases appear. 
In this subsection, we estimate the value of $J = J^{\star}$ 
at which this decoupling sets in. 

First we evaluate the phase diagram at various $J$ in the range of $ 0.03 <
J < 0.3$ by the minimization of $G_{2RSB}$.  
We find that the three glass phases exist at $J \leq 0.1$ while only one glass phase exists at $J \geq 0.2$. 
This means $0.1 < J^{\star} < 0.2$. 
However we can not estimate $J^{\star}$ more accurately by this procedure. 
Our numerical minimization becomes unstable at $0.1 < J < 0.2$ 
because the free energy differences between $G_{1RSB}$ and $G_{2RSB}$ become small.  

\begin{figure}
\centering \includegraphics[width=8.5cm]{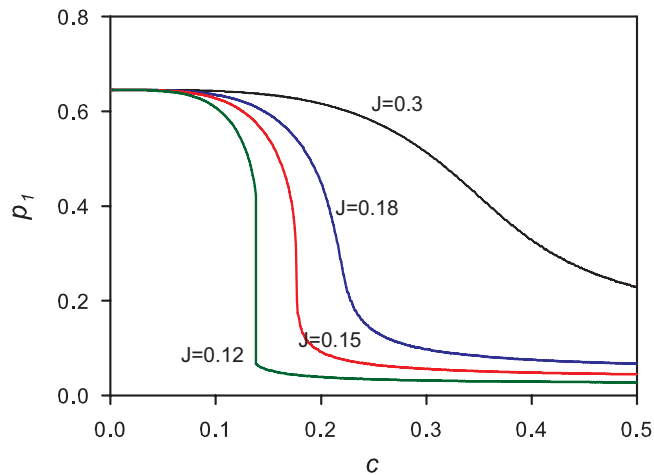} \caption{\label{fig6}
The overlap of weak spins $p_1$ against $c$ on the 1RSB thermodynamic transition line $T_K^{(1)}(c)$ 
at various values of $J$. The glass-glass transition point appears at $J \approx 0.15$.}
\end{figure}

In the course of the evaluations of the phase diagrams described above, 
we find that whenever there exist the three glass phases, 
there also exists the glass-glass transition point ($T^{\star}$, $c^{\star}$) 
from the 1RSB(1) to the 1RSB(2) glass phase on the line $T_K^{(1)}(c)$, see Figure~\ref{fig4}. 
Here, we estimate $J^{\star}$ assuming that $J^{\star}$ is identical to the value of $J$ 
just below which the glass-glass transition point appears on the line $T_K^{(1)}(c)$. 
We evaluate the overlaps along the line $T_K^{(1)}(c)$,  
and seek for the discontinuous jump of the overlaps as a function of $c$,  
which is the sign of the glass-glass transition. 
This analysis is numerically easier than the full evaluation of the phase diagram 
because it requires the numerical minimization only of $G_{1RSB}$. 
In Figure~\ref{fig6}, we show the $c$ dependence of the overlap
of weak spins $p_1$ on the line $T_K^{(1)}(c)$, at several values of $J$. 
At $J=0.3$, $p_1$ decreases smoothly with $c$ and there is no glass-glass transition point.  
The decrease of $p_1$ becomes sharper with decreasing $J$, and becomes discontinuous just below $J \approx 0.15$. 
From this calculation, we estimate $J^{\star}  \approx 0.15$.

\subsection{Thermodynamic quantities}

\begin{figure}
\centering
\includegraphics[width=14cm]{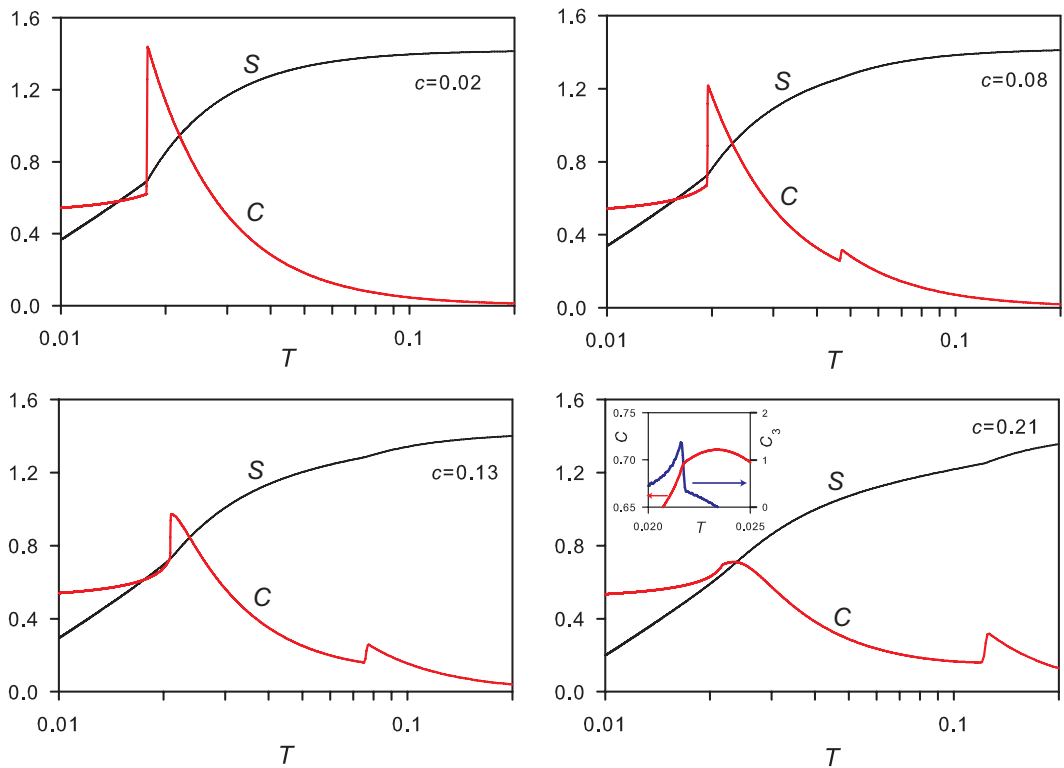}
\caption{\label{fig7} 
Temperature dependence of the entropy $S$ and the heat capacity $C$ 
at $c=0.02$ (upper left), 0.08 (upper right), 0.13 (lower left) and 0.21 (lower right) at $J=0.3$. 
Inset of the lower right panel shows the heat capacity $C$ and the third order derivative of the free energy $C_3$  
at around the continuous 2RSB thermodynamic transition temperature $T_K^{(2c)}$. 
}
\end{figure}

In this subsection, we discuss the nature of the various glass phases 
of the model in terms of the thermodynamic quantities.

Figure~\ref{fig7} shows the temperature dependence of the entropy $S = -
\partial F / \partial T$ and the heat capacity $C = - T \partial^2 F /
\partial T^2$ at $J=0.03$ at the same values of $c$ in
Figure~\ref{fig5}.  These quantities are evaluated by numerical
differentiation of the free energy obtained by the minimization of
$G_{2RSB}$.  At $c=0.02$ (Figure~\ref{fig7} upper left), 
the entropy curve bends and the heat capacity
jumps discontinuously at the 1RSB thermodynamic transition temperature $T_K^{(1)}$.  
This is the typical behavior of the 1RSB glass transition.  
On the other hand at $c=0.08$ (Figure~\ref{fig7} upper right), 
the entropy curve bends and the heat
capacity jumps twice at the 1RSB and the 2RSB thermodynamic transition temperatures, 
$T_K^{(1)}$ and $T_K^{(2d)}$.  
As can be seen from the results at $c=0.13$ (Figure~\ref{fig7} lower left) and 0.21 (Figure~\ref{fig7} lower right), 
the heat capacity jump at $T_K^{(2d)}$ becomes weaker with increasing $c$ 
and eventually disappears when the 2RSB transition becomes continuous.  
In order to characterize the thermodynamics of the continuous 2RSB thermodynamic transition, 
we plot the third order derivative of the free energy
$C_3 = T^2 \partial^3 F / \partial T^3$ in the inset of
Figure~\ref{fig7} lower right.  One finds that this quantity
shows the discontinuous jump at the continuous 2RSB transition temperature $T_K^{(2c)}$.  
Thus the continuous 2RSB transition is the third-order thermodynamic transition in nature.
Note that the similar behavior has been observed for the continuous 1RSB transition~\cite{CS1}. 

\section{Discussion}

We found that the two-component PSM has three glass phases at $J \leq
0.15$: the 1RSB(1) glass where both the strong and weak spins are frozen, 
the 1RSB(2) glass where only the strong spins are frozen, and the
2RSB glass where both the strong and weak spins are frozen and the free
energy landscape has the two-step hierarchical structure.  In this
section, we discuss possible connections and implications of these
results to other systems.

\subsection{Connection to the randamly pinned PSM}

In the randomly pinned PSM, a fraction of spins are pinned 
and the dynamics and thermodynamics of remaining mobile spins are considered. 
This model has recently attracted attention partly, 
because it enables 
us to probe the true thermodynamic glass transition without waiting for the system 
to equilibrate, which otherwise takes astronomically long time~\cite{CammarotaJCP, KB, Ozawa}. 
Interestingly, the 2RSB dynamical and thermodynamic transition lines 
(see Figure~\ref{fig4} right) are analogous to 
the glass transition lines of the randomly pinned glass~\cite{CammarotaJCP}. 
In both cases, the overlap discontinuously jumps at the dynamical transition lines 
when the density of the pinned spins (for the randomly pinned PSM) or 
the strong spins (for the two-component PSM) is small. 
But as the densities increase the discontinuities are weakened and 
eventually the transitions become continuous at which the dynamical transition lines terminate. 
The similarity between these two models is natural because, in the two-component PSM,  
the strong spins frozen at higher temperature behaves as the ``randomly pinned spins''
in the sea of the mobile weak spins at lower temperature.  
Indeed we can establish the precise
relation between these two models.  To this end, we focus on the
behaviors of weak spins in the limit of $J \to 0$ while keeping $T/J$ constant. 
In this limit, the overlaps of strong spins become
$q_2 = 1$, $q_1= 1$ and the breaking parameter $m_1 = 0$.
Plugging these limiting values into $G_{2RSB}$, equation~(\ref{gform2}), 
the relevant part for the weak spins becomes
\begin{eqnarray}
\fl
G_{2RSB} \sim  
 (m_2 -1) [ x_1 + x_2 p_2 + x_3 p_2^2 + x_4 p_2^3] 
- m_2 [x_1 + x_2 p_1 + x_3 p_1^2 + x_4 p_1^3 ]  \nonumber \\
\fl
+ \frac{1-c}{2} \Bigl[
\log (1 - p_2) + \frac{p_1}{1 +(m_2-1)p_2 - m_2 p_1} + \frac{1}{m_2} \log \frac{1 +(m_2-1)p_2 - m_2 p_1}{1 - p_2} \Bigr].  
 \label{gform3}
\end{eqnarray}
This free energy is essentially equivalent to the one of the randomly
pinned PSM~\footnote{ Equation (\ref{gform3}) becomes equivalent to
equation~(16) in Ref.~\cite{CammarotaJCP}, after dividing equation~(16) by
$n$, taking carefully $n \to 0$ limit and replacing $p_2$, $p_1$ and
$m_1$ with $q_1$, $q_0$ and $m$, respectively.}.  Thus the phase diagram
of the two-component PSM converges to that of the randomly pinned PSM
in this limit.

\subsection{Connection to the MCT of binary mixtures}

We next discuss the implications of the two-component PSM 
for binary mixtures of large and small particles with disparate size ratio. 
The MCT was recently used to analyze the decoupling of the glass transitions of large and small particles in this model and predicted the existence of four distinct glass phases~\cite{Voigtmann}: 
(i) The ``partially frozen cageing'' glass 
in which only the large particles are arrested due to the cageing effect
amongst the large particles. 
In this phase, the small particles are left mobile and do not qualitatively affect the dynamics of large particles. 
(ii) The ``partially frozen depletion-driven'' glass
in which only the large particles are arrested by a short-ranged but strong attractive interaction induced 
by the depletion effect caused by small particles~\cite{AO, Dijkstra}. 
In both the phases (i) and (ii), only the large particles undergo the glass transition
and the small particles play a role as the background solvent. 
The phase (i) is often called the repulsive glass and 
(ii) is the attractive glass~\cite{Fuchs1,Fuchs2,Zaccarelli1,Foffi}. 
(iii) The ``fully frozen'' glass
in which both the large and small particles are arrested simultaneously. 
Both the large and small particles equally contribute to the formation of the frozen states. 
(iv) The ``torronchino'' glass 
which is a subset of the ``fully frozen'' glass. 
In this phase, however, 
the number of the small particles is much larger than that of the large particles and 
the freezing is driven mainly by the small particles. 
By comparing the glass phases in our model with those of the MCT, 
one finds that the ``partially frozen cageing'' glass corresponds to the 1RSB(2) glass, 
the ``fully frozen'' to the 2RSB, and the ``torronchino'' to
the 1RSB(1).  Because there is no depletion effect in the present model,
there is no phase corresponding to the ``partially frozen depletion-driven''
glass.  At this stage however, one should realize a subtle but important difference 
between the descriptions of the MCT and the replica theory for these phases. 
Specifically, we revealed that the two-step replica symmetry breaking is
needed to describe the 2RSB glass or the ``fully frozen'' glass. 
However the MCT is believed to be a theory of 
the 1RSB dynamical transition~\cite{BB, Cavagna}, 
therefore it cannot intrinsically describe this phase. 

\begin{figure}
\centering
\includegraphics[width=14cm]{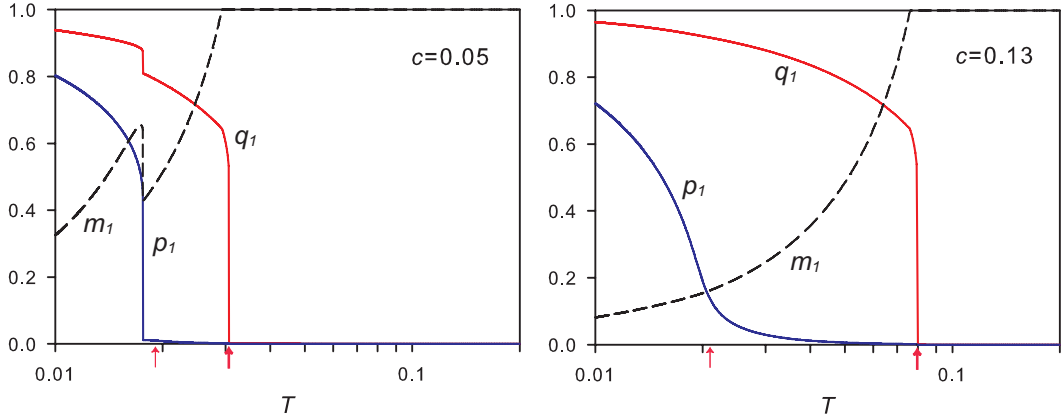}
\caption{\label{fig8}
Temperature dependence of the overlaps $q_1$ and $p_1$ and the breaking parameter $m_1$ 
at $c=0.05$ (left) and 0.13 (right) in the 1RSB solution at $J=0.03$.
The bold and thin red arrows indicate the 1RSB and the 2RSB dynamical transition temperatures, 
$T_d^{(1)}$ and $T_d^{(2)}$, respectively.
The 1RSB solution captures a trace of the 2RSB dynamical transition at $c=0.05$, 
while does not at $c=0.13$. } 
\end{figure}

In order to consider the validity of the prediction of the MCT for the 2RSB glass phase, 
it is useful to see how the
1RSB solution behaves in the 2RSB glass phase~\footnote{
More precisely, the MCT solution corresponds to the 1RSB solution optimized 
with leaving $m_1 = 1$. 
We also performed this calculation and 
verified that the results discussed below are qualitatively unchanged.}. 
In Figure~\ref{fig8}, we
plot the temperature dependence of the overlaps of the 1RSB solution
at $J=0.03$.  At $c=0.05$ (Figure~\ref{fig8} left), 
$p_1$ and $q_1$ jump not only at the 1RSB dynamical transition temperature $T_d^{(1)}$ 
but also at around the 2RSB dynamical transition temperature $T_d^{(2)}$.  
This means that though the 1RSB solution is incorrect in the 2RSB glass region, 
it captures a signature of the transition into the 2RSB glass phase to a certain extent. 
At $c > 0.08$, however, 
we do not find any signature of the 2RSB dynamical transition in the 1RSB solution. 
Indeed at $c=0.13$ (Figure~\ref{fig8} right), 
$q_1$ and $p_1$ increase only smoothly with decreasing temperature in the glass phase, 
while the 2RSB solution predicts the discontinuous 2RSB dynamical transition at $T_d^{(2)}$
(Figure~\ref{fig5} lower left). 
In summary, the 1RSB solution can not correctly describe the 2RSB glass phase
although it can capture a trace of the 2RSB transition for a certain range of parameters. 
This suggests that the applicability of the MCT 
to describe the decoupling of the glass transitions 
in binary mixtures with disparate size ratio may be questioned.

\section{Conclusions}

In this work, we have introduced and studied a two-component version of the $p$-spin
spherical model. 
The model is composed of strongly interacting spins (strong spins) 
and weakly interacting spins (weak spins), 
which mimic the glass forming binary mixtures of large and small particles 
with disparate size ratio.  
We have found that when the
strengths of the interactions of the weak and strong spins are not widely separated,
the model has only one glass phase.  This glass phase is the frozen
state of both the strong and weak spins and is described by the conventional
1RSB solution.  On the other hand when the strengths of the interactions are
well separated, the model exhibits the decoupling of the glass transitions of the weak and strong spins and, as a result, there appear the three distinct glass phases. 
We referred to them as the 1RSB(1), the 1RSB(2), and the 2RSB glass phases. 
The 1RSB(1) glass phase appears in the region 
where the number fraction of the strong spins is very small. 
This glass phase is the frozen
state of both the strong and weak spins, although
the transition into this phase is driven mainly by the freezing of the weak spins. 
The 1RSB(2) glass phase appears in the region 
where the number fraction of the strong spins is large. 
In this glass phase, only the strong spins are frozen while the weak spins are left mobile.   
By cooling the 1RSB(2) glass further, 
the 2RSB glass phase is obtained, in which the weak spins are also frozen. 
The 2RSB glass phase is characterized by the two-step hierarchical structure 
of the free energy landscape.  
The 2RSB glass transition becomes 
ether discontinuous or continuous depending on the number fraction of the strong spins.  
The discontinuous 2RSB thermodynamic transition is accompanied with the jump of the
second order derivative of the free energy, namely the heat capacity. 
On the other hand, for the continuous 2RSB thermodynamic transition, 
the heat capacity changes continuously while 
the third order derivative of the free energy jumps discontinuously. 
Based on the results, we have discussed the connection of the present model 
to the randomly pinned PSM. 
The phase diagram of the present model appears 
to be similar to that of the randomly pinned PSM. 
We have analytically showed that the free energy of the two-component PSM becomes
exactly identical to that of the randomly pinned PSM 
in the small limit of the ratio between 
the strengths of the interactions of the weak and strong spins. 
We have also discussed the implications of the present results 
for the MCT for binary mixtures of large and small particles with disparate size ratio. 
We have found that the 1RSB solution can not correctly describe the 2RSB glass phase
although it can capture a trace of the 2RSB transitions for a certain range of parameters, 
which may leave questionable the applicability of the MCT 
to describe the decoupling of the glass transitions 
in binary mixtures with disparate size ratio. 
Regarding this point, it is interesting to extend the replicated
liquid state theory~\cite{MP, PZ} to allow the 2RSB
ansatz~\cite{EXACT3} to describe the decoupling of the glass transitions 
in binary mixtures with disparate size ratio. 
Study along this direction is under way~\cite{HS2RSB}.

\ack
We thank K. Miyazaki for his careful reading of the manuscript and 
constructive comments. We also thank H. Yoshino for his helpful discussions. 
HI acknowledges the JSPS Core-to-Core program and Program for Leading Graduate Schools 
``Integrative Graduate Education and Research in Green Natural Sciences'', MEXT, Japan, 
and JSPS KAKENHI No. 24340098, 25103005 and 25000002. 
AI acknowledges JSPS KAKENHI No. 26887021.

\appendix
\section{Perturbation analysis of the continuous transition}
\setcounter{section}{1}

In this appendix, we construct the perturbative theory around the 1RSB
ansatz which allows us to calculate the continuous 2RSB thermodynamic transition temperature, $T_K^{(2c)}$. 
Here, we expand the saddle point equations about the differences between
the 1RSB and 2RSB order parameters, $q_2-q_1$ and $p_2-p_1$, and derive
the convenient equations to evaluate $T_K^{(2c)}$. 
To this end, the most convenient staring point is
\begin{eqnarray}
 \frac{2}{(m_2-1)c}\pdif{G_{2RSB}}{q_2}-\frac{2}{(m_1-m_2)(1-c)}\pdif{G_{2RSB}}{q_1}
  = 0,\new
 \frac{2}{(m_2-1)c}\pdif{G_{2RSB}}{q_2}-\frac{2}{(m_1-m_2)(1-c)}\pdif{G_{2RSB}}{q_1}
  = 0.\label{114354_8Dec15}
\end{eqnarray}
After the some manipulations, equations (\ref{114354_8Dec15}) can be rewritten as
\begin{eqnarray}
 \frac{1}{m_2}\left(\frac{1}{1-q_2^{\alpha}}-\frac{1}{1-(1-m_2)q_2^\alpha-m_2q_1^\alpha}\right)
  = M_\alpha,\ \alpha \in \{q,p\},\label{134729_8Dec15}
\end{eqnarray}
where $q_i^q=q_i$ and $q_i^p=p_i$. The kernels, $M_q$ and $M_p$, are
defined as
\begin{eqnarray}
\fl
 M_q = \frac{3}{2T^2}\left\{
c^2(q_2^2-q_1^2)+2J^2c(1-c)(q_2p_2-q_1p_1)+J^2(1-c)^2(p_2^2-p_1^2)\right\},\new
\fl
  M_p = \frac{3}{2T^2}\left\{
J^2c^2(q_2^2-q_1^2)+2J^2c(1-c)(q_2p_2-q_1p_1)+J^2(1-c)^2(p_2^2-p_1^2)\right\}.\label{142622_8Dec15}
\end{eqnarray}
Substituting $q_2^\alpha=q_\alpha$ and $q_1^\alpha=q_\alpha-\delta
q_\alpha$ into equations (\ref{114354_8Dec15}) and expanding $\delta
q_\alpha=\varepsilon q_\alpha^{(1)}+\varepsilon^2
q_\alpha^{(2)}+O(\varepsilon^3)$, one obtains the perturbative series
for $\varepsilon$.  Below, we show that the first order term of
$\varepsilon$ decides the transition temperature and the second order
provides the value of $m_2$ at the transition temperature.

For the first order of the perturbative expansion of 
equation (\ref{134729_8Dec15}) about $\varepsilon$, we obtain
\begin{eqnarray}
\sum_{\beta} M_{\alpha,\beta}q_\beta^{(1)}
 = \frac{1}{(1-q_\alpha)^2}q_\alpha^{(1)},\label{142724_8Dec15}
\end{eqnarray}
where we have defined the auxiliary matrix as
\begin{eqnarray}
 M_{\alpha,\beta}
  = -\left.\pdif{M_{\alpha}}{q_1^\beta}\right|_{\{q_2^\gamma=q_1^\gamma=q^\gamma\}}.
\end{eqnarray}
The necessary condition that the equation (\ref{142724_8Dec15}) has the non-zero
solution is
\begin{eqnarray}
 M_{q,p}M_{p,q} = \left(M_{q,q}-\frac{1}{(1-q)^2}\right)
  \left(M_{p,p}-\frac{1}{(1-p)^2}\right).\label{143846_8Dec15}
\end{eqnarray}
This is the closed equation for $q$, $p$ and the temperature
$T$.  Substituting the 1RSB result for $q$ and $p$, we can solve
equation (\ref{143846_8Dec15}) for $T$ and obtain $T_K^{(2c)}$.  As shown in
Figure~\ref{fig5}, at $T_K^{(2c)}$, the $m_2$ changes
discontinuously from $1$ to some positive value smaller than $1$.  This
value of $m_2$ can be obtained by the second order term of
$\varepsilon$:
\begin{eqnarray}
 \frac{1}{(1-q_\alpha)^2}q_\alpha^{(2)} - \frac{m_2}{(1-q_\alpha)^3}(q_\alpha^{(1)})^2
  = \sum_{\alpha\beta}M_{\alpha,\beta}q_\beta^{(2)} + \sum_{\alpha\beta}M_{\alpha,\beta\gamma}
  q_{\beta}^{(1)}q_\gamma^{(1)},\label{150931_8Dec15}
\end{eqnarray}
where we have defined
\begin{eqnarray}
 M_{a,\beta\gamma} = \left.\frac{1}{2}\pdif{^2M_{\alpha}}{q_1^{\beta}\partial q_1^\gamma}\right|_{\{q_2^\gamma=q_1^\gamma=q^\gamma\}}.
\end{eqnarray}
Note that equation~(\ref{150931_8Dec15}) depends on $q_{\alpha}^{(2)}$, the
value of which is undecided at present.  To remove the terms which
contains $q_{\alpha}^{(2)}$ from equation~(\ref{150931_8Dec15}), inspired by
the perturbative analysis of the MCT\cite{Gotze}, we introduce the left
eigen vector, $l_\alpha$, which satisfies
\begin{eqnarray}
 \sum_{\beta}M_{\alpha,\beta}l_{\beta} = \frac{1}{(1-q_\alpha)^2}l_{\alpha}.\label{103945_26Dec15}
\end{eqnarray}
Using this, $q_\alpha^{(1)}$ can be expressed as
\begin{eqnarray}
q_\alpha^{(1)}=gl_\alpha,\label{184347_8Dec15}
\end{eqnarray}
where $g$ is a constant.
Also, we introduce the right eigen vector by
\begin{eqnarray}
  \sum_{\alpha}r_{\alpha}M_{\alpha,\beta} = \frac{1}{(1-q_\beta)^2}r_{\beta}.\label{103954_26Dec15}
\end{eqnarray}
Multiplying $\sum_{\alpha}r_\alpha$ from
the left of equation (\ref{150931_8Dec15}) and using equation (\ref{184347_8Dec15}), we
finally reach the compact formula for $m_2$:
\begin{eqnarray}
 m_2 = - \frac{\sum_{\alpha\beta\gamma}r_{\alpha}M_{\alpha,\beta\gamma}l_{\beta}l_\gamma}{\sum_{\alpha}r_\alpha l_\alpha^2(1-q_\alpha)^{-3}}.\label{161540_8Dec15}
\end{eqnarray}
From this expression, it is clear that the value of $m_2$ is independent
from the normalization constants of the eigen vectors,
equation~(\ref{103945_26Dec15}) and equation~(\ref{103954_26Dec15}).  The right hand
side of equation~(\ref{161540_8Dec15}) is the function of $q$, $p$, and $T$.  
Substituting the 1RSB results into $q$ and $p$ 
and $T_K^{(2c)}$ calculated by equation (\ref{143846_8Dec15}) into $T$, we obtain the value of $m_2$.

\begin{figure}
\centering
\includegraphics[width=8.5cm]{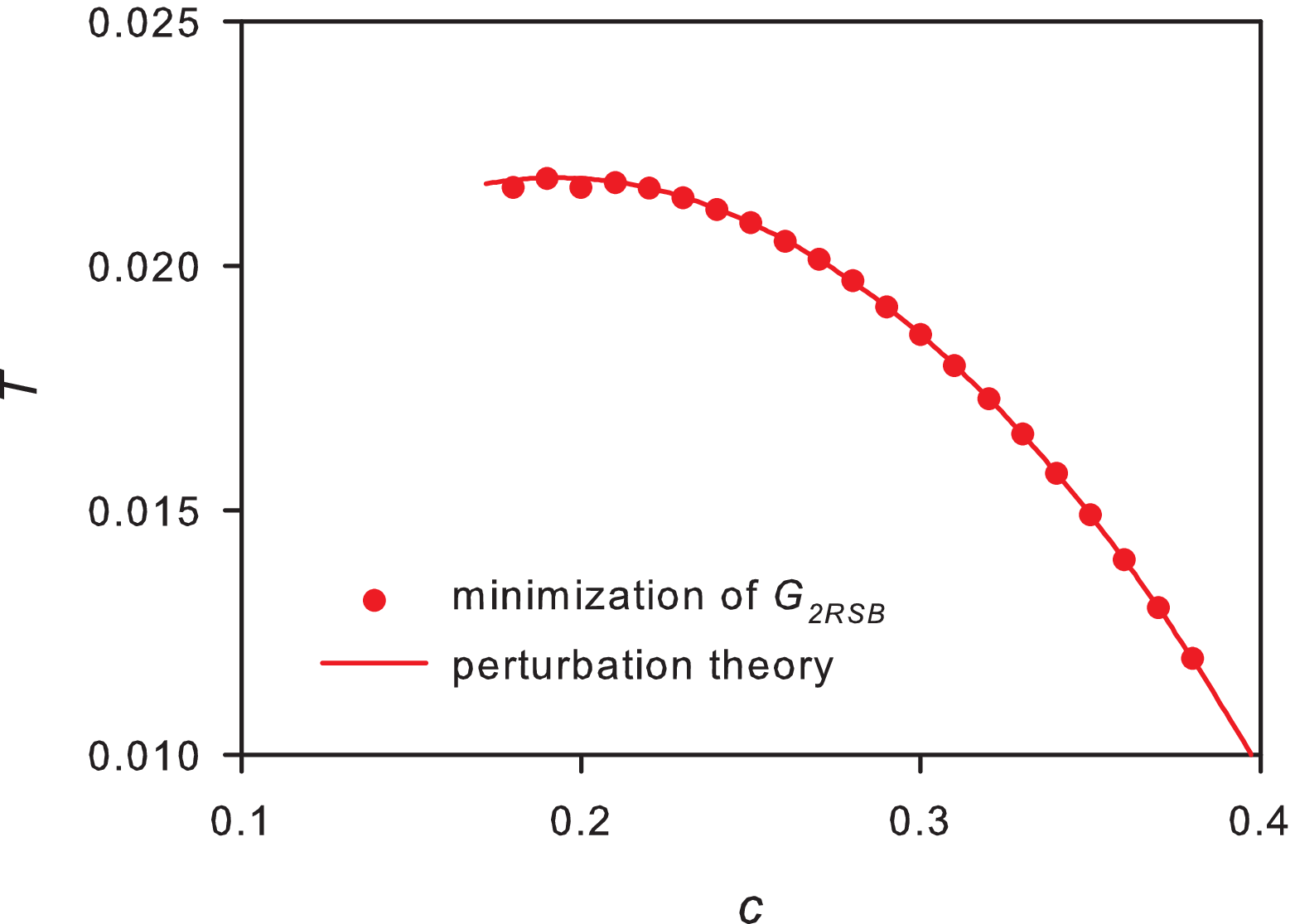}
\caption{\label{figa1} The continuous 2RSB transition temperatures $T_K^{(2c)}$ 
calculated by the minimization of $G_{2RSB}$ (symbols) and equation (\ref{143846_8Dec15}) (line) at $J=0.03$.} 
\end{figure}
An advantage of the formalism constructed above is
that one can evaluate $T_K^{(2c)}$ and the value of
$m_2$ at the transition point with only the information about the 1RSB
result. This enables a more precise investigation of the phase
behavior than that of the full numerical minimization of $G_{2RSB}$.  
In Figure~\ref{figa1}, we compare $T_K^{(2c)}$ 
calculated by equation (\ref{143846_8Dec15}) with that calculated by the minimization of $G_{2RSB}$ (as in Figure.~\ref{fig4}). 
They are almost identical. 

\begin{figure}
\centering
\includegraphics[width=8.5cm]{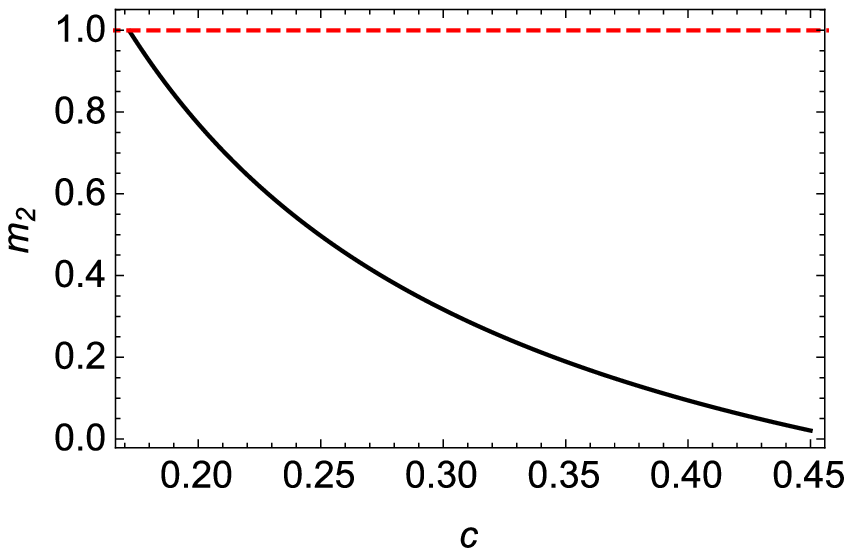}
\caption{\label{figa2}The value of
$m_2$ on the continuous 2RSB transition line  $T_K^{(2c)}$ at $J=0.03$} 
\end{figure}
To determine the critical point at which the continuous transition
ceases to exist and the transition becomes discontinuous, one should
observe the value of 
{$m_2(c)$}.  In Figure~\ref{figa2},
we show the $c$ dependence of 
{$m_2(c)$ calculated by
equation~(\ref{161540_8Dec15})} on 
{$T_K^{(2c)}(c)$}.
{The
value of} $m_2$ increases with decreasing $c$ and reaches $m_2=1$ at the
critical point {$c_c$= 0.17174, where
$T_K^{(2c)}(c_c)=T_c=0.021667$}.
Note that $m_2=1$ is a signal of the
discontinuous transition, therefore it is natural to guess that at
$c=c_c$, 
{the continuous 2RSB
thermodynamic transition line, $T_K^{(2c)}(c)$,} is connected to
{the discontinuous 2RSB
thermodynamic and dynamical transition lines, $T_d^{(2)}(c)$ and
$T_K^{(2d)}(c)$}.  This assumption is indeed correct, see
{Figure~\ref{fig4}}.

\end{document}